\documentclass[prb,showpacs,floatfix,superscriptaddress,reprint]{revtex4-2}
\usepackage{graphicx}
\usepackage{braket}
\usepackage[intlimits]{amsmath}
\newcommand{\ci}{\textsl{i}}
\newcommand{\w}{\omega}
\usepackage{color}

\newcommand{\mt}[1]{\mathcal{#1}}

\newcommand{\affilBIU}{Department of Physics, Jack and Pearl Resnick Institute, Bar-Ilan University, Ramat-Gan 52900, Israel}
\newcommand{\affilAU}{Department of Physics, Ariel University, Ariel 40700, Israel}
\begin{document}

\title{
%Probing Quasi-Particle Signatures: Entanglement Entropy in Excited Many-Particle States\\
%\AK{Visualizing how quasi-particles smear out: entanglement entropy in excited many-particle states}\\
The Smearing of Quasi-Particles: Signatures in the Entanglement Entropy of Excited Many-Particle Systems}

\author{Jagannath Sutradhar}
\affiliation{\affilBIU}
\affiliation{\affilAU}
\author{Jonathan Ruhman}
\affiliation{\affilBIU}
\author{Avraham Klein}
\affiliation{\affilAU}
\author{Dimitri Gutman}
\affiliation{\affilBIU}
\author{Richard Berkovits}
\affiliation{\affilBIU}

\begin{abstract}
 The entanglement spectrum serves as a powerful tool for probing the structure and dynamics of quantum many-body systems, revealing key information about symmetry, topology, and excitations. While the entanglement entropy (EE) of ground states typically follows an area law, highly excited states obey a volume law, leading to a striking contrast in their scaling behavior. In this paper, we investigate the crossover between these two regimes, focusing on the role of quasi-particles (QPs) in mediating this transition. By analyzing the energy dependence of EE in various many-body systems, we explore how the presence of long-lived QPs influences the entanglement structure of excited states. We present numerical results for spinless fermions, a spin chain near a many-body localization transition, and the Sachdev-Ye-Kitaev (SYK) model, which lacks a conventional QP description. Our findings are complemented by a theoretical model based on Fermi liquid theory, providing insight into the interaction-dependent scaling of EE and its consistency with numerical simulations. We find that a hallmark of QPs is a linear dependence of the eigenstate EE on energy, which breaks down at high energy and in the limit of strong interaction. The slope of this linear dependence reflects the QP weight, which reduces with interaction strength.
\end{abstract}

\maketitle

\section{Introduction}

The spectral properties of entanglement are a powerful tool for analyzing various states of matter, providing deep insights into the complex quantum behavior of many-body systems. For example, the entanglement spectrum of a system's ground state reveals critical information about the system's symmetries, topological classifications, and low-energy excitations~\cite{pollmann10,calabrese2004entanglement}. Beyond static characteristics, these spectral features are 
useful
%essential 
for exploring the dynamics of many-body systems far from equilibrium~\cite{kim2013ballistic}, such as  many-body localized systems~\cite{bardarson12,serbyn13,abanin21,PouranvariEPJ024}. 

The von Neumann entanglement entropy (EE) is the simplest measure of entanglement in many-body quantum systems, quantifying the entanglement between two regions of a system. EE is governed by two fundamental scaling laws: the area law and the volume law \cite{amico2008,eisert2010}. The area law characterizes the entanglement in ground states, where EE typically scales with the boundary area of the smaller region rather than its volume. In contrast, the volume law applies to highly excited states, where EE scales with the volume of the smaller region. As a result, the EE of excited states is parametrically larger than that of the ground state.

Highly excited-states EE have recently attracted significant interest \cite{Bianchi2022}. However, the crossover from the ground state to excited states, as a function of excitation energy or excitation number, remains a challenging and less explored problem \cite{swingle13,Bhattacharya2013,Vidmar2017,Vidmar2018,Castro-Alvaredo2018,miao21}.

To address this, a crossover function for EE was proposed based on the equivalence between the reduced density matrix (RDM) of a typical excited state and that of a grand canonical ensemble with a temperature corresponding to the excitation energy \cite{swingle13,miao21}. For critical one-dimensional systems described by conformal field theory, this crossover function exhibits a linear dependence on the temperature of the grand canonical ensemble. However, in the Sachdev-Ye-Kitaev (SYK) model, where an analytical calculation of EE as a function of energy is available \cite{Liu2018,Huang2019}, no such linear crossover behavior is observed.

An important question in the study of quantum many-body systems is the nature of emergent integrability. When present, it allows for an effective description of many-body dynamics through quasi-particles (QPs)—long-lived excitations whose properties closely resemble those of free particles. By QPs, we refer to any single-particle excitation with an asymptotically diverging lifetime, particularly in the low-energy limit. This broad definition encompasses not only fermionic Landau QPs but also bosonic QPs, such as phonons and magnons.

Given the significant role that QPs play in understanding the low-energy behavior of many-particle systems, it is natural to consider their relevance in explaining the crossover between area and  volume law of the EE. Indeed, the presence or absence of QPs provides concrete predictions for the EE of the ground state. In gapped systems, the area law for ground-state EE strictly holds~\cite{hastings2007area}, while in gapless quantum systems, the existence of QP excitations introduces logarithmic corrections to the EE~\cite{calabrese2004entanglement,gioev2006entanglement}. These predictions are essentially a result of counting: since QPs have well defined number statistics, they provide a way to ``count'' single-particle product states both in the many-body ground state and in excited states.

Interestingly, emergent non-interacting particles are also a universal property  of the \emph{highest-energy}  states of a system with a bounded spectrum. This is because the unitarity of wavefunctions limits the bandwidth of any renormalizations due to virtual transitions. Thus, deviations of EE from the QP scaling serves as a direct window into the evolution of a system from a noninteracting to strongly-interacting one.

In this paper, we explore the influence of QPs on the entanglement properties of many-body excited eigenstates. We conduct a numerical analysis of the energy dependence of the eigenstates' EE across three distinct many-body systems, selected to represent a range of QP behaviors.
The first model is spinless fermions with nearest-neighbor interactions on a 1D lattice. The non-interacting limit is a prime example of a system where the excitations are well described QPs at all energies. For the weakly interacting case the low energy behavior is described by a Luttinger liquid with bosonic QPs. When interactions are strong, QPs are destroyed, and the system may become unstable, leading to the formation of ordered states like charge and spin density waves. The next model we consider is the Imbrie model: A spin chain which allegedly undergoes a many-body localization transition \cite{AGKL97,basko06,gornyi05,imbrie16a, PouranvariPRB2021}. Thus, low-lying excitations are localized while higher ones are extended. These properties should be clearly  manifested in the EE. We expect that as long as the excited states are localized the EE does not depend strongly on the energy, while in the metallic regime we are back to a typical QP dependence of the EE on energy.
The last model we consider is in a sense the furthest from the QP picture, namely the complex-fermion Sachdev-Ye-Kitaev (CSYK) model\cite{sachdev93,maldacena99,you17,li17}. For this case we expect to see no linear dependence of the EE on energy, even at low energies \cite{Liu2018,Huang2019}.

 In the two cases discussed above, where QPs are present, the low-energy physics is described by Luttinger liquid theory~\cite{giamarchi2004quantum}. Conformal invariance allows to compute the entanglement scaling of low-lying eigenstates, the dependence on temperature  and track entanglement dynamics following a quench~\cite{calabrese2004entanglement}. A key prediction is that the contribution of these QPs to EE, or equivalently, the entropy they carry, remains independent of the interaction strength. The only influence of the  interaction in this regard is the renormalization of the LL  velocity, which can be absorbed into the inverse temperature, $\beta$. Thus, the slope should scale with the inverse many-body energy bandwidth. Nevertheless, we find that the asymptotic scaling of the eigenstate entanglement entropy (EE) with energy depends on the interaction strength in a way that cannot be fully explained by a simple velocity renormalization. Specifically, the slope of the EE as a function of energy is reduced more strongly with increasing interaction than such a renormalization would predict.

To explain this deviation, 
we employ a toy model where this effect is observed; a one-dimensional Fermi liquid theory, which is a paradigmatic example of  emergent low-energy QPs. We compute the dependence of the EE on interaction strength.  
The crucial effect of interaction in this case is the reduction of the QP weight, though which the entropy per QP is reduced. Namely, we assume that the reduced weight is transferred to localized states that do not contribute. It is important to note that the weight is merely a tool to control the entropy per QP and should not be confused with the actual fermionic QP weight in the one-dimensional systems we consider, which is obviously zero. 
Indeed, using this toy model we find better agreement with the numerical results. Namely, the linear dependence of EE on energy is reproduced, with an interaction dependent slope.

The rest of this paper is organized as follows. In Sec. \ref{s2}, we introduce the models used in this paper for studying of entanglement entropy (EE) across different many-body systems. These include the 1D spinless fermion model, the transverse field Imbrie model, and the CSYK model, each chosen for their distinct quasi-particle behavior. In Sec. \ref{s3}, we outline the methodology for calculating EE using exact diagonalization, describing how the entanglement is quantified for different energy states. Sec. \ref{s4} presents the numerical results for the EE of the clean fermion lattice, highlighting the effects of interaction strength and system size. It also covers the disorder-averaged behavior in the Imbrie model and the unique EE characteristics of the CSYK model. In Sec. \ref{s5}, we develop a theoretical framework based on Fermi liquid theory to capture the quasi-particle contributions to EE, examining how the quasi-particle weight influences entropy behavior as a function of temperature or energy. Finally, Sec. \ref{s6} concludes with a discussion of the implications of our results for understanding many-body localization, quantum chaos, and quasi-particle breakdown in strongly interacting systems.

\section{Models}
\label{s2}

To investigate the role of quasi-particles in determining the dependence of EE on the excitation energy we focus on three models exhibiting different QP behaviors. The first model, corresponding to spinless fermions with nearest-neighbor interactions on a 1D lattice, is given by:
\begin{eqnarray} \label{xxz}
H = 
\displaystyle \sum_{i=1}^{L} \epsilon_i {\hat c}^{\dagger}_{i}{\hat c}_{i}
-t \displaystyle \sum_{j=i}^{L-1}({\hat c}^{\dagger}_{i}{\hat c}_{i+1} + h.c.)
\\ \nonumber
+ U \displaystyle \sum_{j=i}^{L-1}({\hat c}^{\dagger}_{i}{\hat c}_{i} - \frac{1}{2})
({\hat c}^{\dagger}_{i+1}{\hat c}_{i+1} - \frac{1}{2}),
\end{eqnarray}
Here, the on-site energy $\epsilon_i = 0$, indicating a clean system. ${\hat c}_i^{\dagger}$ is the creation operator of a spinless particle at site $i$, $L$ is the number of sites of the lattice, and $t$ is the hopping matrix element set as $t = 1$. The nearest-neighbor interaction strength is given by $U$. The number of electrons $N = L/2 + 1$ corresponds to one electron above half-filling to avoid pathologies associated with half-filling. The size of the Hilbert space is $M = \binom{L}{N}$.

The second model is the transverse magnetic field spin chain with $L$ spins, sometimes referred to as the Imbrie model \cite{imbrie16,imbrie16a}, which is given by:

\begin{eqnarray} \label{imbrie}
  \hat H=\sum_i^{L} h_i \hat S_i^z +  \sum_i \gamma_i \hat S_i^x+
  \sum_i^{L-1} J_{i} \hat S_i^z\hat S_{i+1}^z,
\end{eqnarray}
where $h_i$ is a random magnetic field in the $\hat z$ direction on site $i$, drawn from a box distribution between $-W/2$ and $W/2$, $\gamma_i = 1$, $S_i^a$ is the spin on site $i$ in direction $\hat a$, and $J_{i}$ are the nearest-neighbor spin-spin interactions drawn from a box distribution in the range $0.8$ to $1.2$ \cite{abanin21,berkovits22}. The size of the Hilbert space is $M = 2^L$.

The third model is the complex spinless fermions formulation of the Sachdev-Ye-Kitaev model \cite{you17,li17,berkovits23}
corresponding to:
%model
\begin{eqnarray} \label{syk}
  \hat  H= 
  \sum_{i>j>k>l}^{L} V_{i,j,k,l} \hat c_i^\dag \hat c_j^\dag \hat c_k \hat c_l 
%\\ \nonumber
%\hat  H= \frac{1}{2}\sum_{i,j} J_{i,j} \hat \chi_i^\dag \hat \chi_j +
%\sum_{i,j,k,l} \frac{1}{4!} J_{i,j,k,l}
%\hat \chi_i^\dag \hat \chi_j^\dag \hat \chi_k \hat \chi_l 
\end{eqnarray}
where there are $L$ sites all coupled to each other by a complex interaction, $V_{i,j,k,l}$, where the real and imaginary components are independently drawn from a box distribution between $-L^{-3/2}/2 \ldots L^{-3/2}/2$. The number of fermions is conserved, and we considered the $N = L/2$ sector. The size of the Hilbert space is $M = \binom{L}{N}$.
%for even $L$ and $N = (L+1)/2$ sector for odd $L$.

\section{Entanglement Entropy Calculations}
\label{s3}

For all models using exact diagonalization, we obtain the eigenvalues $\varepsilon_n$ and eigenvectors $|\Psi_n\rangle$ for all $M$ states of the system.

The EE of the $n$-th state $|\Psi_n\rangle$ in a sample partitioned into two sections, A and B, each of length $L_A=L_B=L/2$, is given by

\begin{eqnarray}
S^n_{A}=-{\rm Tr} \rho^n_{A} \ln \rho^n_{A},
\label{sa}
\end{eqnarray}
where the reduced density matrix 
$\rho^n_{A}={\rm Tr}_{B}|\Psi_n\rangle\langle \Psi_n |$.

Thus, using the calculated eigenvectors $|\Psi_n\rangle$ one can construct the reduced density matrix, diagonalize it and using the eigenvalues of $\rho^n_{A}$, calculate 
$S^n_{A}$ for each eigenvector of the Hamiltonian.

\section{Numerical Results}
\label{s4}

\subsection{Fermions in a clean lattice}

For 1D spinless, very weakly-interacting ($U = 0.01$) fermions (Eq. (\ref{xxz})), the entanglement entropy (EE), $S^n_A$, was calculated for many-particle states in the range $1 \leq n \leq M/2$ (covering the energy range from the ground state to the middle of the many-body band) for three system sizes: $L = 14, 16, 18$ with $N = 8, 9, 10$, respectively. This weak interaction is introduced because, in the non-interacting case, many accidental degeneracies occur, leading to ambiguity in defining the n-th eigenstate and, consequently, in determining the EE.

\begin{figure}
  \includegraphics[width=\hsize,clip, trim = 25 25 125 75]{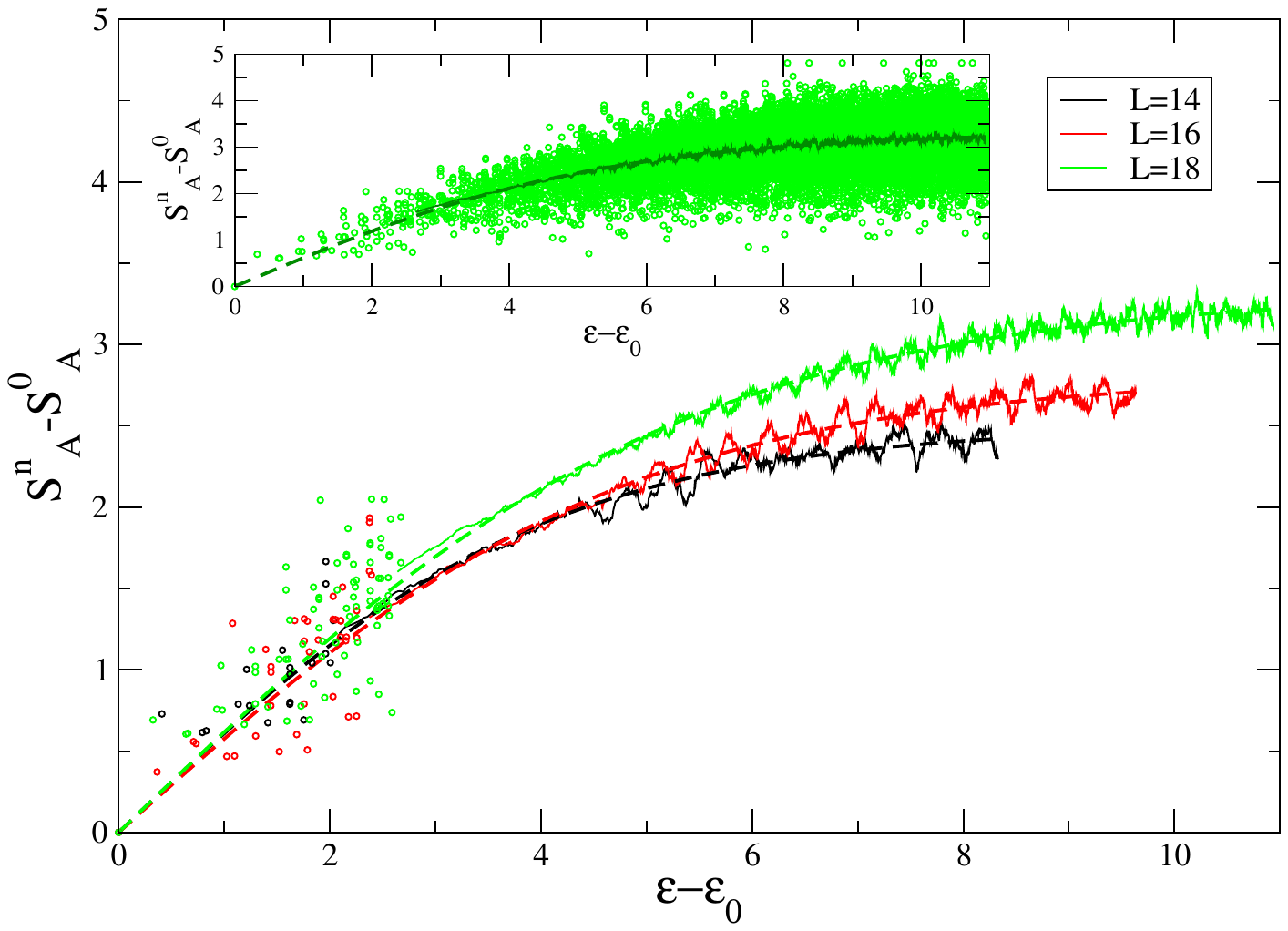}
\caption{\label{fig1}
The excess entanglement entropy (EE) above the ground-state value, ${S^n}_A - {S^0}_A$, between the two halves of the sample is plotted for the $n$-th many-particle state as a function of the many-particle energy above the ground state, $\varepsilon_n - \varepsilon_0$, for three system sizes, $L = 14, 16, 18$, with weak interaction $U = 0.01$. Each system is populated by one electron above half-filling, $N = L/2 + 1$. The continuous curves represent local averages $\langle {S^n}_A \rangle_P$ with $P = 50, 100, 200$ for $L = 14, 16, 18$, respectively. In the inset, the EE values for all states for $L = 18$ are shown as symbols. To reduce visual clutter in the main plot, only the local averages $\langle {S^n}_A \rangle_P$ are displayed for $n > P/2$. The dashed curve represents a fit to the form $C \tanh((\varepsilon_n - \varepsilon_0)/\zeta)$, with fitting parameters $C = 2.50, 2.81, 3.33$ and $\zeta = 4.02, 4.82, 5.33$ for $L = 14, 16, 18$, respectively. 
}
\end{figure}

The results are presented in Fig. \ref{fig1}. In the inset, we plot the energy of the excited state above the ground state, $\varepsilon_n - \varepsilon_0$, versus the excess EE above the ground-state EE, $S^n_A - S^0_A$, for $L = 18$. Significant state-to-state fluctuations in the EE are evident. The fact that EE fluctuation in the energy basis are large compared to the EE fluctuations in a random basis has been seen also for spin systems \cite{Bianchi2019}. Although the fluctuations are large, a general trend can still be observed.  This trend is highlighted by examining the local average $\langle S^n_A \rangle_P$, where $\langle \ldots \rangle_{P}$ denotes the average over states between $n - P/2$ and $n + P/2$ (and is thus only defined for $n > P/2$), represented by the continuous curve in the figure.

To capture the behavior of the EE at all energies, we fit the following function $S^n_A - S^0_A=C \tanh((\varepsilon_n-\varepsilon_0)/\zeta)$, where $C, \zeta$ are fitting parameters. The logic behind this form is that it captures the behavior of the EE both at small and large energies. As can be seen in Fig. \ref{fig1}, the fit is quite good with $C = 2.50, 2.81 , 3.33$, and $\zeta = 4.02, 4.82, 5.33$ for $L = 14, 16, 18$, respectively. Extrapolating to excitations energies close to the ground state results in a linear ${S^n}_A - {S^0}_A=(C/\zeta)(\varepsilon_n-\varepsilon_0)$, where $C/\zeta = 0.62,0.58,0.62$ for the three sizes. Thus, the EE for different system sizes seems to have similar slope at small energies.

\begin{figure}
\includegraphics[width=8.5cm,height=!]{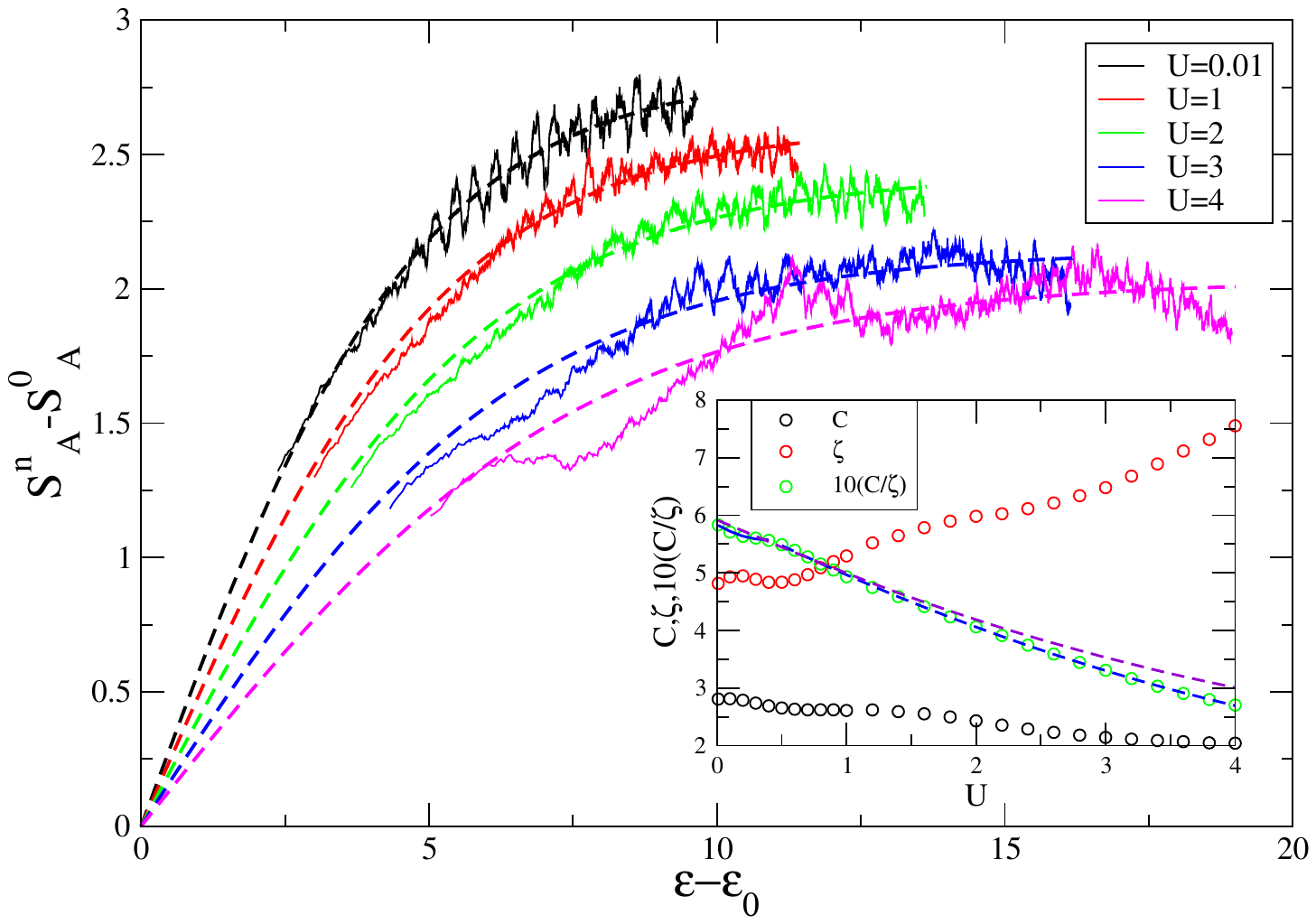}
\caption{\label{fig2}
The entanglement entropy (EE) is plotted for different strength of electron-electron interactions, $U$. The local average EE, $\langle {S^n}_A \rangle_{P=100} - \langle {S^0}_A \rangle_{P=100}$, is shown versus the many-particle energy $\varepsilon_n - \varepsilon_0$, corresponding to the continuous curves. The dashed curves represent fits to the form $C \tanh((\varepsilon_n - \varepsilon_0)/\zeta)$.
Inset: The parameters $C$ and $\zeta$ are plotted for $U$ values in the range $0.01$ to $4$. The ratio $C/\zeta$ (scaled by a factor of $10$) extrapolates the slope at low energies and exhibits a weak dependence on $U$ for small interactions ($U < 0.5$), as indicated by the continuous blue curve. For larger $U$, this ratio decreases approximately linearly. The purple dashed curve corresponds to the inverse bandwidth (multiplied by a constant).
}
\end{figure}

We now examine the role of interactions in a system with $L = 16$ and $N = 9$. As shown in Fig. \ref{fig2}, increasing the strength of the nearest-neighbor electron-electron interaction, $U$, suppresses the eigenstate entanglement entropy (EE) while significantly broadening the energy band as function of $U$. The local average of the EE, $\langle {S^n}A \rangle_{P=100}$, is notably lower compared to the weakly interacting case ($U = 0.01$). To analyze the behavior, we fit the EE to $C \tanh((\varepsilon_n - \varepsilon_0)/\zeta)$. Here, $\zeta$ characterizes the onset of EE saturation, while the slope near the ground state is given by $C/\zeta$. The values of $C$, $\zeta$, and $C/\zeta$ are shown in the inset of Fig. \ref{fig2}.

We observe that both $C$ and $\zeta$ decrease for weak interactions ($U < 0.5$). For stronger interactions, however, $C$ continues to decrease, while $\zeta$ reverses direction and increases. As previously noted, the many-particle bandwidth expands, creating two opposing effects on $\zeta$: a decrease due to the earlier onset of EE saturation and an increase due to the broadening of the energy band. This interplay leads to a minimum in $\zeta$ at $U \sim 0.5$, which also manifests as a shift in the behavior of the slope $C/\zeta$—from a weak dependence on $U$ at low values to an almost linear dependence at higher values of $U$.

In the context of conformal field theory (CFT), the EE depends on the excitation velocity, which in turn is influenced by the interaction strength. A natural explanation for the observed dependence of the linear slope in the EE–interaction strength relation — captured by $C/\zeta$ - is that it reflects the variation of the excitation velocity with interaction strength. In CFT, the velocity is inversely proportional to the many-body energy bandwidth. Therefore, comparing $C/\zeta$ to the inverse bandwidth, as shown in the inset of Fig.\ref{fig2}, should reveal agreement.

However, as the interaction strength $U$ increases, the two curves begin to diverge. This deviation indicates that the dependence of EE on interaction strength is not fully described by CFT. We attribute this discrepancy to contributions from high-energy excitations, where fermionic quasiparticles are expected to emerge. This point will be discussed further in Sec.\ref{s5}.

\subsection{Imbrie model}

For any disordered model, such as the Imbrie model (Eq. \ref{imbrie}), the local averaging used for clean systems can be replaced by averaging over different realizations of disorder, known as ensemble averaging. Here, we have averaged over $100$ different realizations of disorder for each value of disorder strength $W$. Unlike clean systems, there is a qualitative change in the physical properties of the states as energy increases, specifically many-body localization (MBL). For the ground state EE, it has been shown that in the localized regime the entanglement saturates once the system size is larger than the localization length \cite{berkovits12}. Assuming a critical energy, $\varepsilon_c$, one expects that for $\varepsilon<\varepsilon_c$ the entanglement is capped by the energy dependent localiztion length $\xi(E)$. On the other hand, for $\varepsilon>\varepsilon_c$, where the system is in the metallic regime, the EE should increase linearly as long as the state is not to close to the middle of the band. Thus, we expect that the EE in the mettalic regime, i.e., for $\varepsilon_n>\varepsilon_c$, may be described by $\langle {S^n}_A \rangle - \langle {S^0}_A \rangle=-D+C \tanh((\varepsilon_n - \varepsilon_0)/\zeta)$, where $D$ quantifies the energy at which the metallic behavior commences. 

\begin{figure}
  \includegraphics[width=8.5cm,height=!]{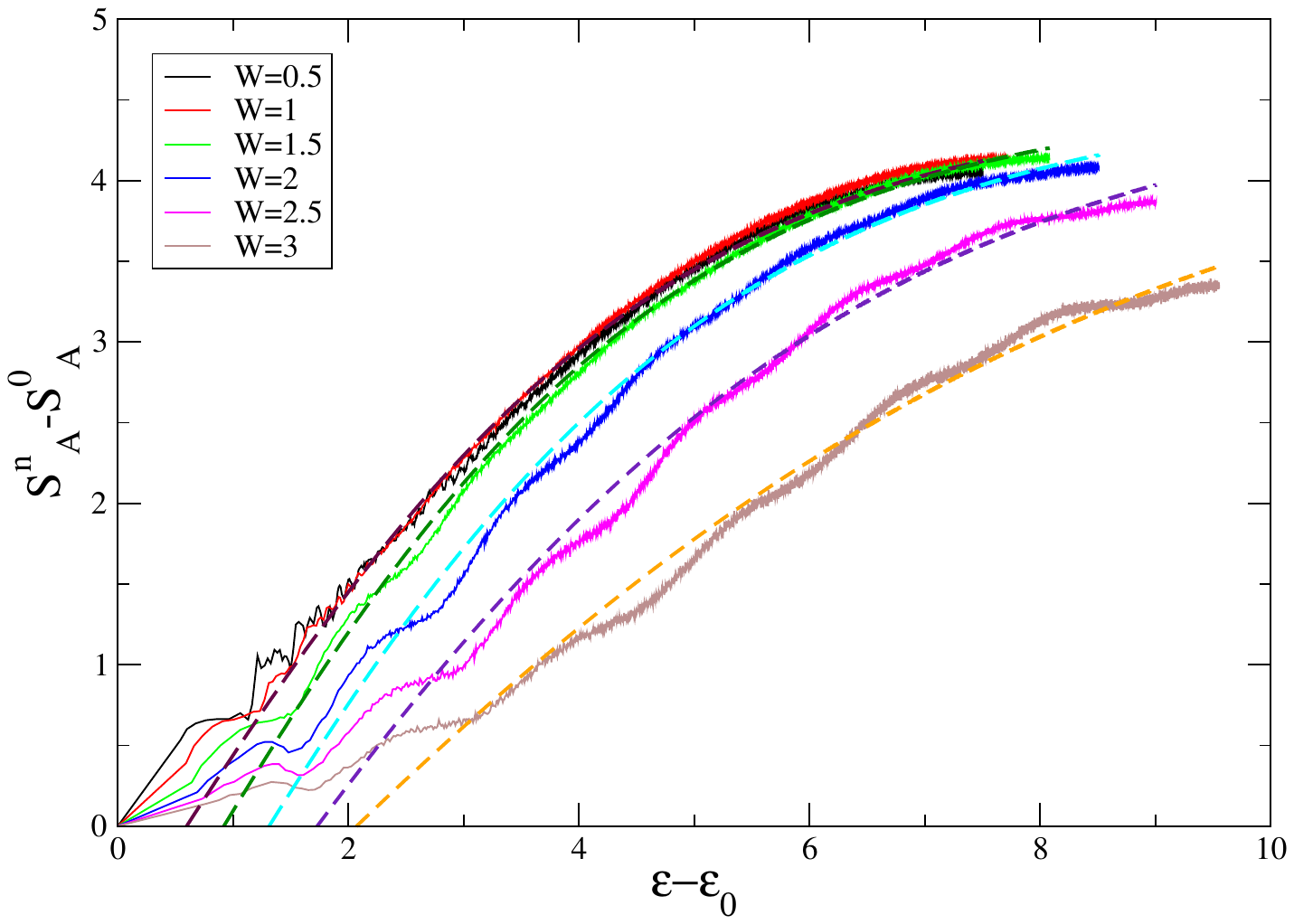}
\caption{\label{fig4}
Averaged EE, $\langle S^n_{A} \rangle$ beyond the ground-state EE $\langle S^0_{A} \rangle$ vs. the many-particle state excitation $\varepsilon_n-\varepsilon_0$, averaged over $100$ realizations of disordered spin chains of length $L=14$ and different strength of disorder $W$. A fit of the EE to $\langle {S^n}_A \rangle - \langle {S^0}_A \rangle=-D+C \tanh((\varepsilon_n - \varepsilon_0)/\zeta)$,
with values of $D,C$ and $\zeta$ given in table \ref{table1} corresponds to the dashed curves.}
\end{figure}

Indeed, examining Fig. \ref{fig4}, one can see that for weak disorder ($W<1$), the metallic behavior begins close to the bottom of the band (small values of $D$, see Table \ref{table1}) and saturates towards the middle of the band. Thus, these realizations are metallic even close to the ground state. For stronger disorder, EE increases slowly at low energies and an onset of the metallic behavior occurs at higher energies commensurate with the picture of a crossover at a critical value of energy. Examining the slope $C/\zeta$ (see Table \ref{table1}) for larger disorder, $W>2$, shows that it becomes smaller, as might be expected. Thus, EE provides a window into the crossover from localized to extended behavior as a function of the excitation energy. In order to clearly characterize the crossover, a finite size scaling procedure which takes into account the localization length dependence on energy, akin to the one proposed in Ref. \cite{berkovits12}, should be constructed and tested. This will be discussed in future work. 

\begin{table}
    \centering
    \begin{tabular}{|c|c|c|c|c|}
    \hline
        $W$ & $D$ & $C$ & $\zeta$ & $C/\zeta$\\
        \hline
         0.5,1& 0.67 & 5.17 & 4.59 & 1.13\\
         1.5& 1.12 & 5.64 & 4.58 & 1.23\\
         2& 1.58 & 6.13 & 4.98 & 1.23\\
         2.5& 1.75 & 6.36 & 6.12 & 1.04 \\
         3& 1.48 & 6.05 & 8.32 & 0.73\\
         \hline
    \end{tabular}
    \caption{Values of the fit parameters $D,C$ and $\zeta$ to $\langle {S^n}_A \rangle - \langle {S^0}_A \rangle=-D+C \tanh((\varepsilon_n - \varepsilon_0)/\zeta)$ proposed in the text and shown in Fig. \ref{fig4}}
    \label{table1}
\end{table}

\subsection{Sachdev-Ye-Kitaev (CSYK) model}

The Complex Sachdev-Ye-Kitaev model is considered a quintessential example of an interacting many-body system that exhibits complete chaotic behavior, lacking a quasi-particle description. This model has gained significant attention due to its unique properties and profound implications for our understanding of quantum chaos and many-body physics \cite{sachdev93,maldacena99,you17,li17,berkovits23}. The EE dependence on the excitation energy has been studied in a couple of publications \cite{Liu2018,Huang2019}. The EE at the middle of the many-particle band is maximal and given by $S^{max}_A=N(\ln(2)-s_0)$, where $N$ is the number of fermions and $s_0=0$ as estimated in \cite{Liu2018} or  $s_0=1/16$ \cite{Huang2019}.

\begin{figure}
  \includegraphics[width=8.5cm,height=!]{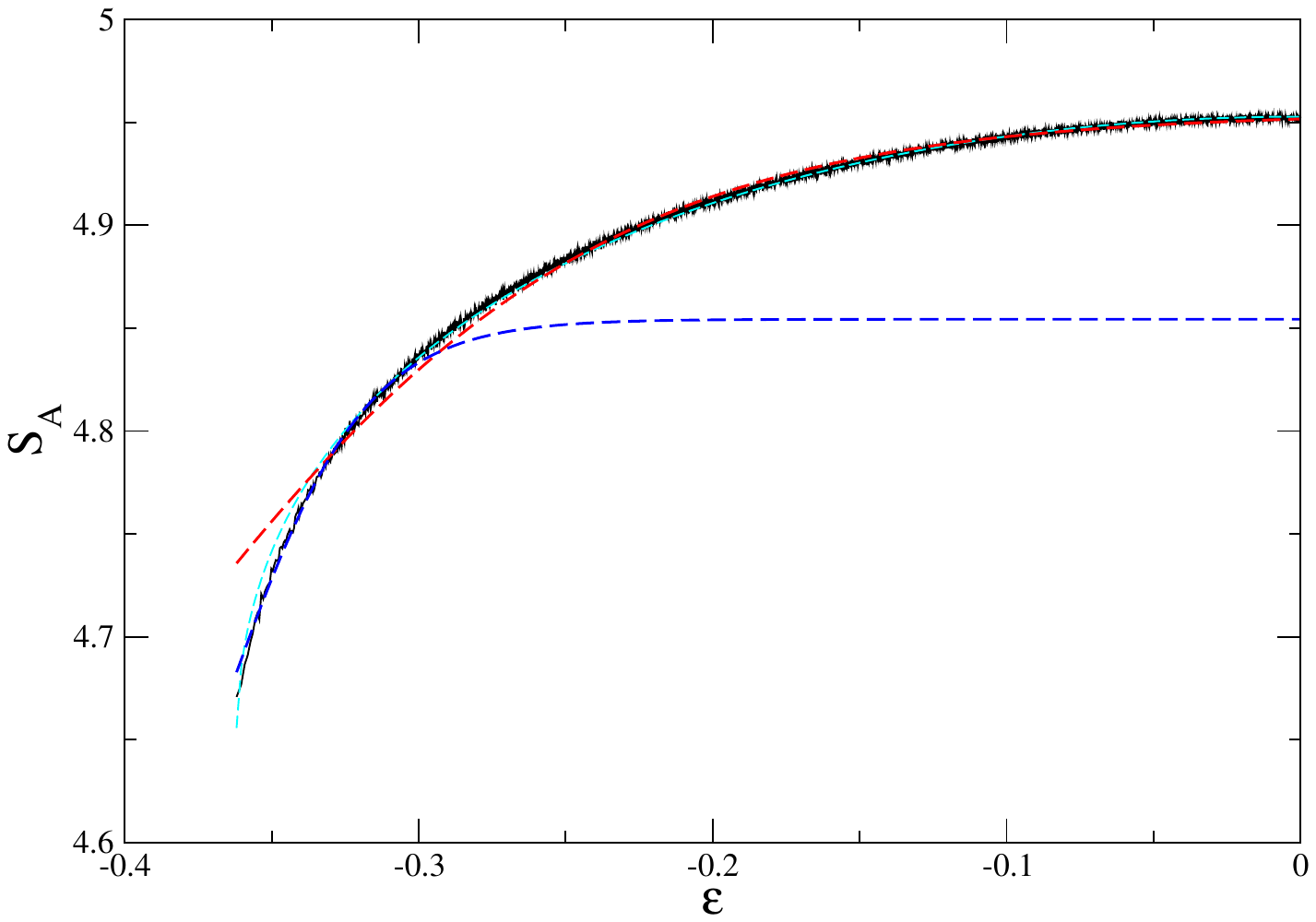}
\caption{\label{fig5}
EE, $\langle S^n_{A} \rangle$, vs. the ground-state energy $\varepsilon$, averaged over $100$ realizations of the Sachdev-Ye-Kitaev (CSYK) model with $L=16$ sites at half-filling. A fit to $S_A(\varepsilon_n) = S^{max}_A - (N/s_1)\arcsin^2(s_2 \varepsilon)$), with $S^{max}_A=N(\ln(2)-s_0)$, where $s_0=1/13.5$, $s_1=65.2$ and $s_2=2.76$, depicted by the green dashed line. Whereas trying to fit the whole range (red dashed) or just states close to the ground state (blue dashed) by $C \tanh((\varepsilon_n - \varepsilon_0)/\zeta)$ are quite poor. 
}
\end{figure}

In Fig. \ref{fig5}, the EE averaged over $100$ realizations for a CSYK model with $L=16$ sites and $N=8$ fermions is depicted as a function of energy. The result for the maximal EE fits the prediction in Refs.~\cite{Liu2018,Huang2019} with $s_0=1/13.5$. Furthermore, as suggested in Ref. \cite{Huang2019} the energy dependence of the EE should follow $S_A(\varepsilon_n) = S^{max}_A - (N/s_1)\arcsin^2(s_2 \varepsilon)$), which with $s_1=65.2$ and $s_2=2.76$, depicted by the dashed green curve, fits very well the numerical result. Unlike for previous models, no successful fits to $\langle {S^n}_A \rangle - \langle {S^0}_A \rangle=C \tanh((\varepsilon_n - \varepsilon_0)/\zeta)$ exists for the CSYK model. Unlike models characterized by a quasi-particle picture, there is no indication of a robust linear dependence, developing into saturation at larger energies. Thus, a typical signature in the EE dependence on energy is absent in a model with no QP's.

\subsection{Theoretical model using Fermi liquid theory}
\label{s5}

To capture the contribution of QPs to the EE discussed earlier, we compute the eigenstate EE of a segment in a one-dimensional Fermi liquid. While Fermi liquid theory breaks down in one spatial dimension, it remains a paradigmatic example of an interacting model with QPs, where the QP weight can be continuously tuned. Moreover, in this case the Fermi liquid theory serves as  a high-energy expansion of the of the single-particle Green's function. 

We evaluate the EE of a Fermi liquid using the correlation matrix $\mt{C}$ 
\begin{equation}
S_A= -\mathrm{Tr}[\mt C\ln\mt C+(1-\mt C)\ln(1-\mt C)],    
\end{equation}
where 
\begin{equation}
\mt{C}_{ij}
= \braket{c_i^{\dagger}(0) c_j(0^-)} 
= G(i-j,0^-). 
\label{Eq:C_G}
\end{equation}
Here, $i,j=1,2,....,L_A$ denote the lattice points within the subsystem $A$, $c_i^{(\dagger)}(\tau)$ is the fermionic operator at $i$-th site at imaginary time $\tau$, and  $G$ is the imaginary time-ordered Green's function. $\braket{..}$ represents a thermal averaging at some temperature $T$.

\begin{figure*}
    \centering
    \includegraphics[width=1.0\linewidth]{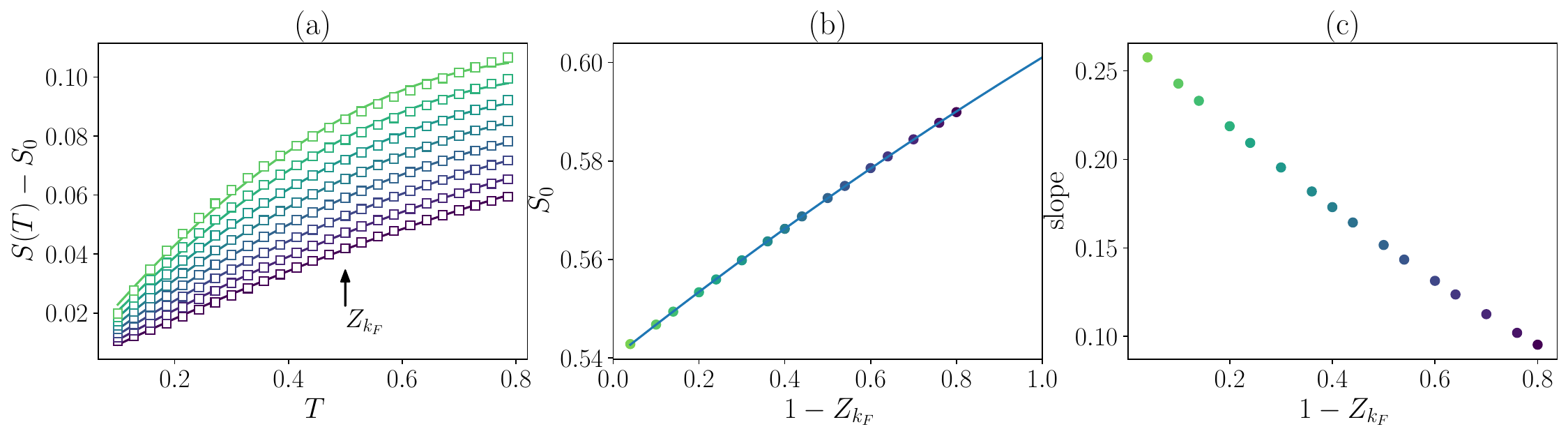}
    \caption{(a) Entropy is shown as a function of $T$ for various QP weight $Z_{k_F}$. The squares represent numerical data points and the solid lines represent fitting to a second-order polynomial. As $Z_{k_F}$ decreases, the dependence of entropy on $T$ becomes weaker. (b) $S_0=S(T=0)$ obtained by extrapolating $S(T)$ to $T=0$ approaches the value $0.6$ as $Z_{k_F}\rightarrow 0$, which represents the contribution from the smooth background excitation in the Green's function. (c) Shows the linear coefficients of the fitting in (a) and it decreases with decreasing $Z_{k_F}$. }
    \label{fig:EE_FL}
\end{figure*}

The Green's function, $G$, consists of two components: a quasiparticle (QP) contribution, weighted by the function $Z_{k}$, and an incoherent background, which we approximate using a Lorentzian function with a width on the order of the bandwidth (for details, see Appendix \ref{sa1}).
The incoherent contribution is required to obey the sum rule of the spectral function in our calculation. 
A finite QP weight, $0 < Z_{k_F} \leq 1$, leads to a discontinuity in the occupation at the Fermi energy in the limit of $T = 0$. In the case of free fermions, $Z_{k_F} = 1$, but as interactions become stronger, $Z_{k_F}$ decreases, indicating a reduction in QP weight.
Thus, to study how the eigenstate EE is affected by interactions we vary the QP weight. 
Supported by the numerical calculations (see Appendix \ref{sa1}) we consider the EE averaged over a small window of energies as equivalent to the EE of a thermal state with the corresponding temperature. 

The entropy per site $S(T)=S_A/L_A$ at half-filling and as a function of $T$ for various QP weights is shown in Fig.~\ref{fig:EE_FL}(a). Here we subtract $S_0=S(T=0)$, obtained by extrapolation of our finite $T$ results. The curves $S(T)-S_0$ vs. $T$ are fitted with a second-order polynomial and show clear linear behavior in the limit $T\to 0$.
The linear slope, plotted in Fig.~\ref{fig:EE_FL}(c), decreases with the QP contribution, $Z$.
This indicates that in the absence of QPs, the dependence of $S$ on $T$ will be very weak, resembling the behavior observed in the CSYK model. Due to numerical constraints, we cannot directly evaluate $S$ at $Z_{k_F}$ and $T\rightarrow 0$.

As mentioned above, a nonzero $S_0$ appears due to the incoherent background. We plot $S_0$ as a function of $Z$ in Fig.~\ref{fig:EE_FL}(c). Its value tends towards $0.6$ as more weight is transferred from the QPs to the incoherent background ($Z\to 0$). This shows that the incoherent part in our phenomenological model contributes to the entropy. However, this clearly depends on the mechanism that leads to loss of QP weight.  For example, close to Mott transition we expect the lost spectral weight to go to localized states which contribute very little to entropy. For this reason, we subtract the zero temperature value. 

In summary, we observe that the existence of QPs leads to a linear in $T$ dependence of the bipartite EE. The slope depends linearly on their weight. Thus, we expect that the stronger the interaction is the smaller the slope is, in agreement with the numerical simulation. 

Our simulations were performed in one spatial dimension. The QP excitations  above the  ground state, if exist,  are captured by a LL, which is very different in nature compared to the QPs we used here, which belong to Fermi liquid theory. 
Namely, these QPs are longitudinal density waves. 
Nonetheless, these  excitations still have a ``weight" and we expect it to reduce with interaction strength. For example, at any intermediate scale there are corrections to the LL coming from irrelevant terms. These are expected to reduce the strength of the  QP contribution to correlations. The stronger the interactions are the larger these corrections are going to be as well.

% However, in principle, in the presence of strong interaction and at $T=0$, the ground state of a system should be a Mott insulator yielding zero EE. The unavoidable smooth function produces the spurious nonzero result, which is why we subtract it.

\section{Discussion}
We examined how quasi-particle excitations influence the bipartite entanglement entropy of eigenstates. We analyzed a range of key models, specifically spinless fermions with nearest-neighbor interactions, the Imbrie model, and the complex fermionic SYK model. We found that when quasi-particles are present, they contribute a linear-in-energy component to the entanglement entropy. The slope of this contribution is related to the quasi-particle weight.

The origin of this dependence can be understood by examining the many-body density of states. When it is governed by quasi-particles, it grows exponentially with energy. In this situation, thermal states, which are a classical ensemble of pure state density matrices made of these eigenstates, behave like the typical eigenstate. To accommodate with thermodynamics the typical state must therefore have an entropy that grows linearly with energy. 
This should be contrasted with systems without quasi-particles, where the many-body density of states no longer grows exponentially with energy.

An interesting question regards the exact calculation of the quasi-particle weight within Luttinger liquid theory instead of Fermi liquid theory. This can be done by computing the overlap between the low-energy counting field and free bosonic excitations associated with density (e.g. a harmonic chain). At any finite scale the non-harmonic terms are finite and thus, also couple to the harmonic excitations reducing the overlap. 

Another important question regards the breakdown of the quasiparticle behavior we observed at large interaction strength. A series of minima in the eigenstate EE emerges at well defined energies that seem to be regularly spaced. It is interesting to understand whether these features are related to the formation of quasi bands and whether they reflect tendency towards ordered states.

\label{s6}
\bibliography{EE}

\begin{thebibliography}{35}%
\makeatletter
\providecommand \@ifxundefined [1]{%
 \@ifx{#1\undefined}
}%
\providecommand \@ifnum [1]{%
 \ifnum #1\expandafter \@firstoftwo
 \else \expandafter \@secondoftwo
 \fi
}%
\providecommand \@ifx [1]{%
 \ifx #1\expandafter \@firstoftwo
 \else \expandafter \@secondoftwo
 \fi
}%
\providecommand \natexlab [1]{#1}%
\providecommand \enquote  [1]{``#1''}%
\providecommand \bibnamefont  [1]{#1}%
\providecommand \bibfnamefont [1]{#1}%
\providecommand \citenamefont [1]{#1}%
\providecommand \href@noop [0]{\@secondoftwo}%
\providecommand \href [0]{\begingroup \@sanitize@url \@href}%
\providecommand \@href[1]{\@@startlink{#1}\@@href}%
\providecommand \@@href[1]{\endgroup#1\@@endlink}%
\providecommand \@sanitize@url [0]{\catcode `\\12\catcode `\$12\catcode
  `\&12\catcode `\#12\catcode `\^12\catcode `\_12\catcode `\%12\relax}%
\providecommand \@@startlink[1]{}%
\providecommand \@@endlink[0]{}%
\providecommand \url  [0]{\begingroup\@sanitize@url \@url }%
\providecommand \@url [1]{\endgroup\@href {#1}{\urlprefix }}%
\providecommand \urlprefix  [0]{URL }%
\providecommand \Eprint [0]{\href }%
\providecommand \doibase [0]{https://doi.org/}%
\providecommand \selectlanguage [0]{\@gobble}%
\providecommand \bibinfo  [0]{\@secondoftwo}%
\providecommand \bibfield  [0]{\@secondoftwo}%
\providecommand \translation [1]{[#1]}%
\providecommand \BibitemOpen [0]{}%
\providecommand \bibitemStop [0]{}%
\providecommand \bibitemNoStop [0]{.\EOS\space}%
\providecommand \EOS [0]{\spacefactor3000\relax}%
\providecommand \BibitemShut  [1]{\csname bibitem#1\endcsname}%
\let\auto@bib@innerbib\@empty
%</preamble>
\bibitem [{\citenamefont {Pollmann}\ \emph {et~al.}(2010)\citenamefont
  {Pollmann}, \citenamefont {Turner}, \citenamefont {Berg},\ and\ \citenamefont
  {Oshikawa}}]{pollmann10}%
  \BibitemOpen
  \bibfield  {author} {\bibinfo {author} {\bibfnamefont {F.}~\bibnamefont
  {Pollmann}}, \bibinfo {author} {\bibfnamefont {A.~M.}\ \bibnamefont
  {Turner}}, \bibinfo {author} {\bibfnamefont {E.}~\bibnamefont {Berg}},\ and\
  \bibinfo {author} {\bibfnamefont {M.}~\bibnamefont {Oshikawa}},\ }\bibfield
  {title} {\bibinfo {title} {Entanglement spectrum of a topological phase in
  one dimension},\ }\href@noop {} {\bibfield  {journal} {\bibinfo  {journal}
  {Physical Review B}\ }\textbf {\bibinfo {volume} {81}},\ \bibinfo {pages}
  {064439} (\bibinfo {year} {2010})}\BibitemShut {NoStop}%
\bibitem [{\citenamefont {Calabrese}\ and\ \citenamefont
  {Cardy}(2004)}]{calabrese2004entanglement}%
  \BibitemOpen
  \bibfield  {author} {\bibinfo {author} {\bibfnamefont {P.}~\bibnamefont
  {Calabrese}}\ and\ \bibinfo {author} {\bibfnamefont {J.}~\bibnamefont
  {Cardy}},\ }\bibfield  {title} {\bibinfo {title} {Entanglement entropy and
  quantum field theory},\ }\href@noop {} {\bibfield  {journal} {\bibinfo
  {journal} {Journal of statistical mechanics: theory and experiment}\ }\textbf
  {\bibinfo {volume} {2004}},\ \bibinfo {pages} {P06002} (\bibinfo {year}
  {2004})}\BibitemShut {NoStop}%
\bibitem [{\citenamefont {Kim}\ and\ \citenamefont
  {Huse}(2013)}]{kim2013ballistic}%
  \BibitemOpen
  \bibfield  {author} {\bibinfo {author} {\bibfnamefont {H.}~\bibnamefont
  {Kim}}\ and\ \bibinfo {author} {\bibfnamefont {D.~A.}\ \bibnamefont {Huse}},\
  }\bibfield  {title} {\bibinfo {title} {Ballistic spreading of entanglement in
  a diffusive nonintegrable system},\ }\href@noop {} {\bibfield  {journal}
  {\bibinfo  {journal} {Physical review letters}\ }\textbf {\bibinfo {volume}
  {111}},\ \bibinfo {pages} {127205} (\bibinfo {year} {2013})}\BibitemShut
  {NoStop}%
\bibitem [{\citenamefont {Bardarson}\ \emph {et~al.}(2012)\citenamefont
  {Bardarson}, \citenamefont {Pollmann},\ and\ \citenamefont
  {Moore}}]{bardarson12}%
  \BibitemOpen
  \bibfield  {author} {\bibinfo {author} {\bibfnamefont {J.~H.}\ \bibnamefont
  {Bardarson}}, \bibinfo {author} {\bibfnamefont {F.}~\bibnamefont
  {Pollmann}},\ and\ \bibinfo {author} {\bibfnamefont {J.~E.}\ \bibnamefont
  {Moore}},\ }\bibfield  {title} {\bibinfo {title} {Unbounded growth of
  entanglement in models of many-body localization},\ }\href@noop {} {\bibfield
   {journal} {\bibinfo  {journal} {Physical review letters}\ }\textbf {\bibinfo
  {volume} {109}},\ \bibinfo {pages} {017202} (\bibinfo {year}
  {2012})}\BibitemShut {NoStop}%
\bibitem [{\citenamefont {Serbyn}\ \emph {et~al.}(2013)\citenamefont {Serbyn},
  \citenamefont {Papi{\'c}},\ and\ \citenamefont {Abanin}}]{serbyn13}%
  \BibitemOpen
  \bibfield  {author} {\bibinfo {author} {\bibfnamefont {M.}~\bibnamefont
  {Serbyn}}, \bibinfo {author} {\bibfnamefont {Z.}~\bibnamefont {Papi{\'c}}},\
  and\ \bibinfo {author} {\bibfnamefont {D.~A.}\ \bibnamefont {Abanin}},\
  }\bibfield  {title} {\bibinfo {title} {Universal slow growth of entanglement
  in interacting strongly disordered systems},\ }\href@noop {} {\bibfield
  {journal} {\bibinfo  {journal} {Physical review letters}\ }\textbf {\bibinfo
  {volume} {110}},\ \bibinfo {pages} {260601} (\bibinfo {year}
  {2013})}\BibitemShut {NoStop}%
\bibitem [{\citenamefont {Abanin}\ \emph {et~al.}(2021)\citenamefont {Abanin},
  \citenamefont {Bardarson}, \citenamefont {De~Tomasi}, \citenamefont
  {Gopalakrishnan}, \citenamefont {Khemani}, \citenamefont {Parameswaran},
  \citenamefont {Pollmann}, \citenamefont {Potter}, \citenamefont {Serbyn},\
  and\ \citenamefont {Vasseur}}]{abanin21}%
  \BibitemOpen
  \bibfield  {author} {\bibinfo {author} {\bibfnamefont {D.}~\bibnamefont
  {Abanin}}, \bibinfo {author} {\bibfnamefont {J.~H.}\ \bibnamefont
  {Bardarson}}, \bibinfo {author} {\bibfnamefont {G.}~\bibnamefont
  {De~Tomasi}}, \bibinfo {author} {\bibfnamefont {S.}~\bibnamefont
  {Gopalakrishnan}}, \bibinfo {author} {\bibfnamefont {V.}~\bibnamefont
  {Khemani}}, \bibinfo {author} {\bibfnamefont {S.}~\bibnamefont
  {Parameswaran}}, \bibinfo {author} {\bibfnamefont {F.}~\bibnamefont
  {Pollmann}}, \bibinfo {author} {\bibfnamefont {A.}~\bibnamefont {Potter}},
  \bibinfo {author} {\bibfnamefont {M.}~\bibnamefont {Serbyn}},\ and\ \bibinfo
  {author} {\bibfnamefont {R.}~\bibnamefont {Vasseur}},\ }\bibfield  {title}
  {\bibinfo {title} {Distinguishing localization from chaos: Challenges in
  finite-size systems},\ }\href@noop {} {\bibfield  {journal} {\bibinfo
  {journal} {Annals of Physics}\ }\textbf {\bibinfo {volume} {427}},\ \bibinfo
  {pages} {168415} (\bibinfo {year} {2021})}\BibitemShut {NoStop}%
\bibitem [{\citenamefont {Pouranvari}(2024)}]{PouranvariEPJ024}%
  \BibitemOpen
  \bibfield  {author} {\bibinfo {author} {\bibfnamefont {M.}~\bibnamefont
  {Pouranvari}},\ }\bibfield  {title} {\bibinfo {title} {Directional
  localization in disordered 2d tight-binding systems: insights from
  single-particle entanglement measures},\ }\href
  {https://doi.org/10.1140/epjb/s10051-024-00824-y} {\bibfield  {journal}
  {\bibinfo  {journal} {The European Physical Journal B}\ }\textbf {\bibinfo
  {volume} {97}},\ \bibinfo {pages} {180} (\bibinfo {year} {2024})}\BibitemShut
  {NoStop}%
\bibitem [{\citenamefont {Amico}\ \emph {et~al.}(2008)\citenamefont {Amico},
  \citenamefont {Fazio}, \citenamefont {Osterloh},\ and\ \citenamefont
  {Vedral}}]{amico2008}%
  \BibitemOpen
  \bibfield  {author} {\bibinfo {author} {\bibfnamefont {L.}~\bibnamefont
  {Amico}}, \bibinfo {author} {\bibfnamefont {R.}~\bibnamefont {Fazio}},
  \bibinfo {author} {\bibfnamefont {A.}~\bibnamefont {Osterloh}},\ and\
  \bibinfo {author} {\bibfnamefont {V.}~\bibnamefont {Vedral}},\ }\bibfield
  {title} {\bibinfo {title} {Entanglement in many-body systems},\ }\href
  {https://doi.org/10.1103/RevModPhys.80.517} {\bibfield  {journal} {\bibinfo
  {journal} {Reviews of Modern Physics}\ }\textbf {\bibinfo {volume} {80}},\
  \bibinfo {pages} {517} (\bibinfo {year} {2008})}\BibitemShut {NoStop}%
\bibitem [{\citenamefont {Eisert}\ \emph {et~al.}(2010)\citenamefont {Eisert},
  \citenamefont {Cramer},\ and\ \citenamefont {Plenio}}]{eisert2010}%
  \BibitemOpen
  \bibfield  {author} {\bibinfo {author} {\bibfnamefont {J.}~\bibnamefont
  {Eisert}}, \bibinfo {author} {\bibfnamefont {M.}~\bibnamefont {Cramer}},\
  and\ \bibinfo {author} {\bibfnamefont {M.~B.}\ \bibnamefont {Plenio}},\
  }\bibfield  {title} {\bibinfo {title} {Area laws for the entanglement entropy
  - a review},\ }\href {https://doi.org/10.1103/RevModPhys.82.277} {\bibfield
  {journal} {\bibinfo  {journal} {Reviews of Modern Physics}\ }\textbf
  {\bibinfo {volume} {82}},\ \bibinfo {pages} {277} (\bibinfo {year}
  {2010})}\BibitemShut {NoStop}%
\bibitem [{\citenamefont {Bianchi}\ \emph {et~al.}(2022)\citenamefont
  {Bianchi}, \citenamefont {Hackl}, \citenamefont {Kieburg}, \citenamefont
  {Rigol},\ and\ \citenamefont {Vidmar}}]{Bianchi2022}%
  \BibitemOpen
  \bibfield  {author} {\bibinfo {author} {\bibfnamefont {E.}~\bibnamefont
  {Bianchi}}, \bibinfo {author} {\bibfnamefont {L.}~\bibnamefont {Hackl}},
  \bibinfo {author} {\bibfnamefont {M.}~\bibnamefont {Kieburg}}, \bibinfo
  {author} {\bibfnamefont {M.}~\bibnamefont {Rigol}},\ and\ \bibinfo {author}
  {\bibfnamefont {L.}~\bibnamefont {Vidmar}},\ }\bibfield  {title} {\bibinfo
  {title} {Volume-law entanglement entropy of typical pure quantum states},\
  }\href {https://doi.org/10.1103/PRXQuantum.3.030201} {\bibfield  {journal}
  {\bibinfo  {journal} {PRX Quantum}\ }\textbf {\bibinfo {volume} {3}},\
  \bibinfo {pages} {030201} (\bibinfo {year} {2022})},\ \bibinfo {note} {and
  references within}\BibitemShut {NoStop}%
\bibitem [{\citenamefont {Swingle}\ and\ \citenamefont
  {Senthil}(2013)}]{swingle13}%
  \BibitemOpen
  \bibfield  {author} {\bibinfo {author} {\bibfnamefont {B.}~\bibnamefont
  {Swingle}}\ and\ \bibinfo {author} {\bibfnamefont {T.}~\bibnamefont
  {Senthil}},\ }\bibfield  {title} {\bibinfo {title} {Universal crossovers
  between entanglement entropy and thermal entropy},\ }\href@noop {} {\bibfield
   {journal} {\bibinfo  {journal} {Physical Review B—Condensed Matter and
  Materials Physics}\ }\textbf {\bibinfo {volume} {87}},\ \bibinfo {pages}
  {045123} (\bibinfo {year} {2013})}\BibitemShut {NoStop}%
\bibitem [{\citenamefont {Bhattacharya}\ \emph {et~al.}(2013)\citenamefont
  {Bhattacharya}, \citenamefont {Nozaki}, \citenamefont {Takayanagi},\ and\
  \citenamefont {Ugajin}}]{Bhattacharya2013}%
  \BibitemOpen
  \bibfield  {author} {\bibinfo {author} {\bibfnamefont {J.}~\bibnamefont
  {Bhattacharya}}, \bibinfo {author} {\bibfnamefont {M.}~\bibnamefont
  {Nozaki}}, \bibinfo {author} {\bibfnamefont {T.}~\bibnamefont {Takayanagi}},\
  and\ \bibinfo {author} {\bibfnamefont {T.}~\bibnamefont {Ugajin}},\
  }\bibfield  {title} {\bibinfo {title} {Thermodynamical property of
  entanglement entropy for excited states},\ }\href
  {https://doi.org/10.1103/PhysRevLett.110.091602} {\bibfield  {journal}
  {\bibinfo  {journal} {Physical Review Letters}\ }\textbf {\bibinfo {volume}
  {110}},\ \bibinfo {pages} {091602} (\bibinfo {year} {2013})}\BibitemShut
  {NoStop}%
\bibitem [{\citenamefont {Vidmar}\ \emph {et~al.}(2017)\citenamefont {Vidmar},
  \citenamefont {Hackl}, \citenamefont {Bianchi},\ and\ \citenamefont
  {Rigol}}]{Vidmar2017}%
  \BibitemOpen
  \bibfield  {author} {\bibinfo {author} {\bibfnamefont {L.}~\bibnamefont
  {Vidmar}}, \bibinfo {author} {\bibfnamefont {L.}~\bibnamefont {Hackl}},
  \bibinfo {author} {\bibfnamefont {E.}~\bibnamefont {Bianchi}},\ and\ \bibinfo
  {author} {\bibfnamefont {M.}~\bibnamefont {Rigol}},\ }\bibfield  {title}
  {\bibinfo {title} {Entanglement entropy of eigenstates of quadratic fermionic
  hamiltonians},\ }\href {https://doi.org/10.1103/PhysRevLett.119.020601}
  {\bibfield  {journal} {\bibinfo  {journal} {Physical Review Letters}\
  }\textbf {\bibinfo {volume} {119}},\ \bibinfo {pages} {020601} (\bibinfo
  {year} {2017})}\BibitemShut {NoStop}%
\bibitem [{\citenamefont {Vidmar}\ \emph {et~al.}(2018)\citenamefont {Vidmar},
  \citenamefont {Santos},\ and\ \citenamefont {Rigol}}]{Vidmar2018}%
  \BibitemOpen
  \bibfield  {author} {\bibinfo {author} {\bibfnamefont {L.}~\bibnamefont
  {Vidmar}}, \bibinfo {author} {\bibfnamefont {L.~F.}\ \bibnamefont {Santos}},\
  and\ \bibinfo {author} {\bibfnamefont {M.}~\bibnamefont {Rigol}},\ }\bibfield
   {title} {\bibinfo {title} {Volume law and quantum criticality in the
  entanglement entropy of excited eigenstates of the quantum ising model},\
  }\href {https://doi.org/10.1103/PhysRevLett.121.220602} {\bibfield  {journal}
  {\bibinfo  {journal} {Physical Review Letters}\ }\textbf {\bibinfo {volume}
  {121}},\ \bibinfo {pages} {220602} (\bibinfo {year} {2018})}\BibitemShut
  {NoStop}%
\bibitem [{\citenamefont {Castro-Alvaredo}\ \emph {et~al.}(2018)\citenamefont
  {Castro-Alvaredo}, \citenamefont {Doyon},\ and\ \citenamefont
  {Yoshimura}}]{Castro-Alvaredo2018}%
  \BibitemOpen
  \bibfield  {author} {\bibinfo {author} {\bibfnamefont {O.~A.}\ \bibnamefont
  {Castro-Alvaredo}}, \bibinfo {author} {\bibfnamefont {B.}~\bibnamefont
  {Doyon}},\ and\ \bibinfo {author} {\bibfnamefont {T.}~\bibnamefont
  {Yoshimura}},\ }\bibfield  {title} {\bibinfo {title} {Entanglement content of
  quasiparticle excitations},\ }\href
  {https://doi.org/10.1103/PhysRevLett.121.170602} {\bibfield  {journal}
  {\bibinfo  {journal} {Physical Review Letters}\ }\textbf {\bibinfo {volume}
  {121}},\ \bibinfo {pages} {170602} (\bibinfo {year} {2018})}\BibitemShut
  {NoStop}%
\bibitem [{\citenamefont {Miao}\ and\ \citenamefont {Barthel}(2021)}]{miao21}%
  \BibitemOpen
  \bibfield  {author} {\bibinfo {author} {\bibfnamefont {Q.}~\bibnamefont
  {Miao}}\ and\ \bibinfo {author} {\bibfnamefont {T.}~\bibnamefont {Barthel}},\
  }\bibfield  {title} {\bibinfo {title} {Eigenstate entanglement: Crossover
  from the ground state to volume laws},\ }\href@noop {} {\bibfield  {journal}
  {\bibinfo  {journal} {Physical Review Letters}\ }\textbf {\bibinfo {volume}
  {127}},\ \bibinfo {pages} {040603} (\bibinfo {year} {2021})}\BibitemShut
  {NoStop}%
\bibitem [{\citenamefont {Chunxiao~Liu}\ and\ \citenamefont
  {Balents}(2018)}]{Liu2018}%
  \BibitemOpen
  \bibfield  {author} {\bibinfo {author} {\bibfnamefont {X.~C.}\ \bibnamefont
  {Chunxiao~Liu}}\ and\ \bibinfo {author} {\bibfnamefont {L.}~\bibnamefont
  {Balents}},\ }\bibfield  {title} {\bibinfo {title} {Quantum entanglement of
  the sachdev-ye-kitaev models},\ }\href
  {https://doi.org/10.1103/PhysRevB.97.245126} {\bibfield  {journal} {\bibinfo
  {journal} {Physical Review B}\ }\textbf {\bibinfo {volume} {97}},\ \bibinfo
  {pages} {245126} (\bibinfo {year} {2018})}\BibitemShut {NoStop}%
\bibitem [{\citenamefont {Huang}\ and\ \citenamefont {Gu}(2019)}]{Huang2019}%
  \BibitemOpen
  \bibfield  {author} {\bibinfo {author} {\bibfnamefont {Y.}~\bibnamefont
  {Huang}}\ and\ \bibinfo {author} {\bibfnamefont {Y.}~\bibnamefont {Gu}},\
  }\bibfield  {title} {\bibinfo {title} {Eigenstate entanglement in the
  sachdev-ye-kitaev model},\ }\href
  {https://doi.org/10.1103/PhysRevD.100.041901} {\bibfield  {journal} {\bibinfo
   {journal} {Physical Review D}\ }\textbf {\bibinfo {volume} {100}},\ \bibinfo
  {pages} {041901(R)} (\bibinfo {year} {2019})}\BibitemShut {NoStop}%
\bibitem [{\citenamefont {Hastings}(2007)}]{hastings2007area}%
  \BibitemOpen
  \bibfield  {author} {\bibinfo {author} {\bibfnamefont {M.~B.}\ \bibnamefont
  {Hastings}},\ }\bibfield  {title} {\bibinfo {title} {An area law for
  one-dimensional quantum systems},\ }\href@noop {} {\bibfield  {journal}
  {\bibinfo  {journal} {Journal of statistical mechanics: theory and
  experiment}\ }\textbf {\bibinfo {volume} {2007}},\ \bibinfo {pages} {P08024}
  (\bibinfo {year} {2007})}\BibitemShut {NoStop}%
\bibitem [{\citenamefont {Gioev}\ and\ \citenamefont
  {Klich}(2006)}]{gioev2006entanglement}%
  \BibitemOpen
  \bibfield  {author} {\bibinfo {author} {\bibfnamefont {D.}~\bibnamefont
  {Gioev}}\ and\ \bibinfo {author} {\bibfnamefont {I.}~\bibnamefont {Klich}},\
  }\bibfield  {title} {\bibinfo {title} {Entanglement entropy of fermions in
  any dimension and the widom conjecture},\ }\href@noop {} {\bibfield
  {journal} {\bibinfo  {journal} {Physical review letters}\ }\textbf {\bibinfo
  {volume} {96}},\ \bibinfo {pages} {100503} (\bibinfo {year}
  {2006})}\BibitemShut {NoStop}%
\bibitem [{\citenamefont {Altshuler}\ \emph {et~al.}(1997)\citenamefont
  {Altshuler}, \citenamefont {Gefen}, \citenamefont {Kamenev},\ and\
  \citenamefont {Levitov}}]{AGKL97}%
  \BibitemOpen
  \bibfield  {author} {\bibinfo {author} {\bibfnamefont {B.~L.}\ \bibnamefont
  {Altshuler}}, \bibinfo {author} {\bibfnamefont {Y.}~\bibnamefont {Gefen}},
  \bibinfo {author} {\bibfnamefont {A.}~\bibnamefont {Kamenev}},\ and\ \bibinfo
  {author} {\bibfnamefont {L.~S.}\ \bibnamefont {Levitov}},\ }\bibfield
  {title} {\bibinfo {title} {Quasiparticle lifetime in a finite system: A
  nonperturbative approach},\ }\href@noop {} {\bibfield  {journal} {\bibinfo
  {journal} {Phys. Rev. Lett.}\ }\textbf {\bibinfo {volume} {78}},\ \bibinfo
  {pages} {2803} (\bibinfo {year} {1997})}\BibitemShut {NoStop}%
\bibitem [{\citenamefont {Basko}\ \emph {et~al.}(2006)\citenamefont {Basko},
  \citenamefont {Aleiner},\ and\ \citenamefont {Altshuler}}]{basko06}%
  \BibitemOpen
  \bibfield  {author} {\bibinfo {author} {\bibfnamefont {D.~M.}\ \bibnamefont
  {Basko}}, \bibinfo {author} {\bibfnamefont {I.~L.}\ \bibnamefont {Aleiner}},\
  and\ \bibinfo {author} {\bibfnamefont {B.~L.}\ \bibnamefont {Altshuler}},\
  }\bibfield  {title} {\bibinfo {title} {Metal–insulator transition in a
  weakly interacting many-electron system with localized single-particle
  states},\ }\href@noop {} {\bibfield  {journal} {\bibinfo  {journal} {Annals
  of Physics}\ }\textbf {\bibinfo {volume} {321}},\ \bibinfo {pages} {1126}
  (\bibinfo {year} {2006})}\BibitemShut {NoStop}%
\bibitem [{\citenamefont {Gornyi}\ \emph {et~al.}(2005)\citenamefont {Gornyi},
  \citenamefont {Mirlin},\ and\ \citenamefont {Polyakov}}]{gornyi05}%
  \BibitemOpen
  \bibfield  {author} {\bibinfo {author} {\bibfnamefont {I.~V.}\ \bibnamefont
  {Gornyi}}, \bibinfo {author} {\bibfnamefont {A.~D.}\ \bibnamefont {Mirlin}},\
  and\ \bibinfo {author} {\bibfnamefont {D.~G.}\ \bibnamefont {Polyakov}},\
  }\bibfield  {title} {\bibinfo {title} {Interacting electrons in disordered
  wires: Anderson localization and low-$t$ transport},\ }\href@noop {}
  {\bibfield  {journal} {\bibinfo  {journal} {Physical Review Letters}\
  }\textbf {\bibinfo {volume} {95}},\ \bibinfo {pages} {206603} (\bibinfo
  {year} {2005})}\BibitemShut {NoStop}%
\bibitem [{\citenamefont {Imbrie}(2016{\natexlab{a}})}]{imbrie16a}%
  \BibitemOpen
  \bibfield  {author} {\bibinfo {author} {\bibfnamefont {J.~Z.}\ \bibnamefont
  {Imbrie}},\ }\bibfield  {title} {\bibinfo {title} {On many-body localization
  for quantum spin chains},\ }\href@noop {} {\bibfield  {journal} {\bibinfo
  {journal} {Journal of Statistical Physics}\ }\textbf {\bibinfo {volume}
  {163}},\ \bibinfo {pages} {998} (\bibinfo {year}
  {2016}{\natexlab{a}})}\BibitemShut {NoStop}%
\bibitem [{\citenamefont {Pouranvari}\ and\ \citenamefont
  {Liou}(2021)}]{PouranvariPRB2021}%
  \BibitemOpen
  \bibfield  {author} {\bibinfo {author} {\bibfnamefont {M.}~\bibnamefont
  {Pouranvari}}\ and\ \bibinfo {author} {\bibfnamefont {S.-F.}\ \bibnamefont
  {Liou}},\ }\bibfield  {title} {\bibinfo {title} {Characterizing many-body
  localization via state sensitivity to boundary conditions},\ }\href
  {https://doi.org/10.1103/PhysRevB.103.035136} {\bibfield  {journal} {\bibinfo
   {journal} {Phys. Rev. B}\ }\textbf {\bibinfo {volume} {103}},\ \bibinfo
  {pages} {035136} (\bibinfo {year} {2021})}\BibitemShut {NoStop}%
\bibitem [{\citenamefont {Sachdev}\ and\ \citenamefont {Ye}(1993)}]{sachdev93}%
  \BibitemOpen
  \bibfield  {author} {\bibinfo {author} {\bibfnamefont {S.}~\bibnamefont
  {Sachdev}}\ and\ \bibinfo {author} {\bibfnamefont {J.}~\bibnamefont {Ye}},\
  }\bibfield  {title} {\bibinfo {title} {Gapless spin-fluid ground state in a
  random quantum heisenberg magnet},\ }\href@noop {} {\bibfield  {journal}
  {\bibinfo  {journal} {Physical review letters}\ }\textbf {\bibinfo {volume}
  {70}},\ \bibinfo {pages} {3339} (\bibinfo {year} {1993})}\BibitemShut
  {NoStop}%
\bibitem [{\citenamefont {Maldacena}(1999)}]{maldacena99}%
  \BibitemOpen
  \bibfield  {author} {\bibinfo {author} {\bibfnamefont {J.}~\bibnamefont
  {Maldacena}},\ }\bibfield  {title} {\bibinfo {title} {The large-n limit of
  superconformal field theories and supergravity},\ }\href@noop {} {\bibfield
  {journal} {\bibinfo  {journal} {International journal of theoretical
  physics}\ }\textbf {\bibinfo {volume} {38}},\ \bibinfo {pages} {1113}
  (\bibinfo {year} {1999})}\BibitemShut {NoStop}%
\bibitem [{\citenamefont {You}\ \emph {et~al.}(2017)\citenamefont {You},
  \citenamefont {Ludwig},\ and\ \citenamefont {Xu}}]{you17}%
  \BibitemOpen
  \bibfield  {author} {\bibinfo {author} {\bibfnamefont {Y.-Z.}\ \bibnamefont
  {You}}, \bibinfo {author} {\bibfnamefont {A.~W.}\ \bibnamefont {Ludwig}},\
  and\ \bibinfo {author} {\bibfnamefont {C.}~\bibnamefont {Xu}},\ }\bibfield
  {title} {\bibinfo {title} {Sachdev-ye-kitaev model and thermalization on the
  boundary of many-body localized fermionic symmetry-protected topological
  states},\ }\href@noop {} {\bibfield  {journal} {\bibinfo  {journal} {Physical
  Review B}\ }\textbf {\bibinfo {volume} {95}},\ \bibinfo {pages} {115150}
  (\bibinfo {year} {2017})}\BibitemShut {NoStop}%
\bibitem [{\citenamefont {Li}\ \emph {et~al.}(2017)\citenamefont {Li},
  \citenamefont {Liu}, \citenamefont {Xin},\ and\ \citenamefont {Zhou}}]{li17}%
  \BibitemOpen
  \bibfield  {author} {\bibinfo {author} {\bibfnamefont {T.}~\bibnamefont
  {Li}}, \bibinfo {author} {\bibfnamefont {J.}~\bibnamefont {Liu}}, \bibinfo
  {author} {\bibfnamefont {Y.}~\bibnamefont {Xin}},\ and\ \bibinfo {author}
  {\bibfnamefont {Y.}~\bibnamefont {Zhou}},\ }\bibfield  {title} {\bibinfo
  {title} {Supersymmetric syk model and random matrix theory},\ }\href@noop {}
  {\bibfield  {journal} {\bibinfo  {journal} {Journal of High Energy Physics}\
  }\textbf {\bibinfo {volume} {2017}},\ \bibinfo {pages} {1} (\bibinfo {year}
  {2017})}\BibitemShut {NoStop}%
\bibitem [{\citenamefont {Giamarchi}(2004)}]{giamarchi2004quantum}%
  \BibitemOpen
  \bibfield  {author} {\bibinfo {author} {\bibfnamefont {T.}~\bibnamefont
  {Giamarchi}},\ }\href@noop {} {\bibinfo {title} {Quantum physics in one
  dimension}} (\bibinfo {year} {2004})\BibitemShut {NoStop}%
\bibitem [{\citenamefont {Imbrie}(2016{\natexlab{b}})}]{imbrie16}%
  \BibitemOpen
  \bibfield  {author} {\bibinfo {author} {\bibfnamefont {J.~Z.}\ \bibnamefont
  {Imbrie}},\ }\bibfield  {title} {\bibinfo {title} {Diagonalization and
  many-body localization for a disordered quantum spin chain},\ }\href@noop {}
  {\bibfield  {journal} {\bibinfo  {journal} {Physical review letters}\
  }\textbf {\bibinfo {volume} {117}},\ \bibinfo {pages} {027201} (\bibinfo
  {year} {2016}{\natexlab{b}})}\BibitemShut {NoStop}%
\bibitem [{\citenamefont {Berkovits}(2022)}]{berkovits22}%
  \BibitemOpen
  \bibfield  {author} {\bibinfo {author} {\bibfnamefont {R.}~\bibnamefont
  {Berkovits}},\ }\bibfield  {title} {\bibinfo {title} {Large-scale behavior of
  energy spectra of the quantum random antiferromagnetic ising chain with mixed
  transverse and longitudinal fields},\ }\href
  {https://doi.org/10.1103/PhysRevB.105.104203} {\bibfield  {journal} {\bibinfo
   {journal} {Phys. Rev. B}\ }\textbf {\bibinfo {volume} {105}},\ \bibinfo
  {pages} {104203} (\bibinfo {year} {2022})}\BibitemShut {NoStop}%
\bibitem [{\citenamefont {Berkovits}(2023)}]{berkovits23}%
  \BibitemOpen
  \bibfield  {author} {\bibinfo {author} {\bibfnamefont {R.}~\bibnamefont
  {Berkovits}},\ }\bibfield  {title} {\bibinfo {title} {Sachdev-ye-kitaev
  model: Non-self-averaging properties of the energy spectrum},\ }\href
  {https://doi.org/10.1103/PhysRevB.107.035141} {\bibfield  {journal} {\bibinfo
   {journal} {Phys. Rev. B}\ }\textbf {\bibinfo {volume} {107}},\ \bibinfo
  {pages} {035141} (\bibinfo {year} {2023})}\BibitemShut {NoStop}%
\bibitem [{\citenamefont {Bianchi}\ and\ \citenamefont
  {Don{\`a}}(2019)}]{Bianchi2019}%
  \BibitemOpen
  \bibfield  {author} {\bibinfo {author} {\bibfnamefont {E.}~\bibnamefont
  {Bianchi}}\ and\ \bibinfo {author} {\bibfnamefont {P.}~\bibnamefont
  {Don{\`a}}},\ }\bibfield  {title} {\bibinfo {title} {Typical entanglement
  entropy in the presence of a center: Page curve and its variance},\ }\href
  {https://doi.org/10.1103/PhysRevD.100.105010} {\bibfield  {journal} {\bibinfo
   {journal} {Physical Review D}\ }\textbf {\bibinfo {volume} {100}},\ \bibinfo
  {pages} {105010} (\bibinfo {year} {2019})}\BibitemShut {NoStop}%
\bibitem [{\citenamefont {Berkovits}(2012)}]{berkovits12}%
  \BibitemOpen
  \bibfield  {author} {\bibinfo {author} {\bibfnamefont {R.}~\bibnamefont
  {Berkovits}},\ }\bibfield  {title} {\bibinfo {title} {Entanglement entropy in
  a one-dimensional disordered interacting system: The role of localization},\
  }\href {https://doi.org/10.1103/PhysRevLett.108.176803} {\bibfield  {journal}
  {\bibinfo  {journal} {Physical Review Letters}\ }\textbf {\bibinfo {volume}
  {108}},\ \bibinfo {pages} {176803} (\bibinfo {year} {2012})}\BibitemShut
  {NoStop}%
\end{thebibliography}%

\newpage

\appendix

\section{Analytic computation of the EE in a Fermi liquid}
\label{sa1}
%{\color{red}JR: Jagannath, please move this to the appendix and refer to it appropriately in the main text.}
The Green's function in Eq.~\eqref{Eq:C_G} in the momentum and Mastubara frequency space is written as
 \begin{equation}
G(i-j,0^-) 
=\dfrac{1}{\beta} \sum_k \sum_n G(k,\w_n) e^{\ci k(i-j)} e^{\ci\w_n 0^-}. 
\end{equation}
Using the spectral density $A(k,\w)$, we express 
\begin{equation}
G(k,\w_n)=\int_{-\infty}^{\infty} \dfrac{d\w}{2\pi}\dfrac{A(k,\w)}{\ci\w_n-\w},
\end{equation}
where $A(k,\w)$ is related to the retarded Green's function as
\begin{equation}
A(k,\w) = -2\mathrm{Im} G^R(k,\w).
\end{equation}
Therefore,
\begin{align}
\mt{C}_{ij}
&=
 \sum_k \int_{-\infty}^{\infty} \dfrac{d\w}{2\pi} e^{\ci k(i-j)} \oint \dfrac{dz}{2\ci\pi} \dfrac{A(k,\w)}{z-\w} n_F(z) \nonumber \\ 
&=
\sum_k e^{\ci k(i-j)} \int_{-\infty}^{\infty} \dfrac{d\w}{2\pi}  A(k,\w) n_F(\w),
\label{Eq:Corr}
\end{align}
where $n_F$ is the Fermi-Dirac distribution arising in the standard complex integration to sum over the Mastubara frequencies. 

The spectral density has two parts in it such that
$$A(k,\w)=A_{QP}(k,\w)+A_{BG}(k,\w),$$
where $A_{QP}(k,\w)$ is the quasiparticle contribution given by 
\begin{equation}
A_{QP}(k,\w) = -2\dfrac{Z_k^2\Gamma(k,\w)}{ (\w- \epsilon_k)^2 + \Gamma^2(k,\w) Z_k^2}.  
\label{Eq:A_QP}    
\end{equation}
Here, $\Gamma(k,\w)$ is the imaginary part of the self-energy, and we assume the real part to be zero, since it only renormalizes the bare dispersion $\epsilon_k$. We approximate $\Gamma(k,\w)= a(\epsilon_k^2+T^2)$ with a proportionality constant $a$ as a free parameter.
\begin{figure}
    \centering \includegraphics[width=1.0\linewidth]{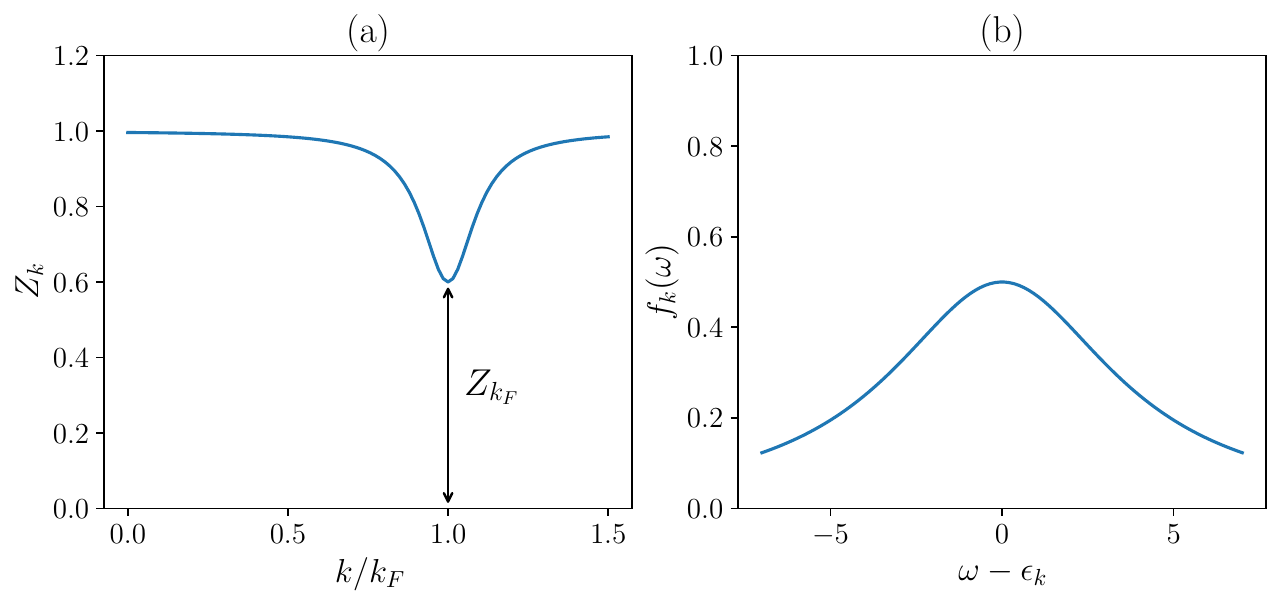}
    \caption{(a) We model $Z_k$ using a Lorentzian dip near the fermi momentum, where $Z_k<1$ due to interaction. The missing $(1-Z_k)$ amount of spectral weight contributes to the smooth particle-hole excitation for each $k$. (b) We model this smooth background using another Lorentz function $f_k(\omega)$ having a width comparable to the bandwidth.}
    \label{fig:model_functions}
\end{figure}

$A_{BG}(k,\w)$ represents the smooth background excitation. 
To keep the sum rule, $\int_{-\infty}^{\infty} A(k,\w)d\w/(2\pi)=1$, intact we approximate $A_{BG}(k,\w)=(1-Z_k)f_k(\w)$, as $\int_{-\infty}^{\infty} A_{QP}(k,\w)d\w/(2\pi)=Z_k$. $f_k(\w)$ is a smooth broad range function of $\w$ with $\int_{-\infty}^{\infty} f_k(\w)d\w/(2\pi)=1$. $f_k(\w)$ is approximated as  $2W/[(\w-\epsilon_k)^2+W^2]$ for $W\gg \epsilon_{k_F}$. For example, Fig.~\ref{fig:model_functions}(b) illustrates $f_k(\w)$ whose center is at $\w=\epsilon_k$ for each $k$.

We model $Z_k=1-2\lambda b_Z/[\epsilon_k^2+b_Z^2] $ with $\lambda$ and $b_Z$ as free parameters, as shown in Fig.~\ref{fig:model_functions}(a) for $b_Z=0.1$ and $\lambda=0.02$. 

\end{document}